\documentstyle[prl,aps]{revtex}
\title{On an Extended PCAC Relationship}
\author{S.Ying\\
\it Department of Physics, Fudan University\\
\it Shanghai, China}
\begin{document}
\draft
\maketitle

\begin{abstract}
{
  We explore a consistent way to extend the partially conserved axial
vector current (PCAC) relationship and corresponding current algebra results
in two strongly correlated directions: 1) towards a search for a set of
systematic rules for the establishment of PCAC related relationships in a
finite low momentum transfer region and for the extrapolation of the momentum
transfer $q^2$ to zero when deriving the low energy PCAC results that can be
compared to experimental data and 2) towards taking into account, besides the
conventional one, the only other possibility of the spontaneous chiral symmetry
breaking, $SU(2)_L\times SU(2)_R \to SU(2)_V$, inside a baryonic system by
a condensation (in the sense to be specified in the paper) of diquarks. The
paper includes investigations of a chiral Ward--Takahashi identity, the
explicit chiral symmetry breaking by a finite current quark mass, the
modification of the PCAC relationship and its consequences. It is shown that
the signals for a hypothetical diquark condensation inside a nucleon and
nucleus is observable in high precision experiments despite the fact that they
may evade most of the current observations. We briefly discuss how diquark
condensation could provide an answer to the question of where about of pions
and quark number in a nucleon and a nucleus, which is raised in explaining
puzzles in polarized pion nucleus scattering, violation of Gottfried sum rule
and EMC effects.
}
\end{abstract}

\pacs{PACS number: 11.30.Hv, 11.30.Qc, 11.30.Rd, 11.40.Ha, 40.20.Dh, 14.65.Bt}
\tighten

\section{Introduction}

    The partially conserved axial vector current (PCAC) relationship and
corresponding current algebra results are in conformity with experimental data
within a few percent. This good agreement between theory and experiments is
interpreted as due to an underlying approximate chiral $SU(2)_L\times SU(2)_R$
symmetry, which is explicitly broken down by up and down current quark masses
of a few MeV much smaller than the hadronic mass scale of 1 GeV. The lightest
hadronic particle pions are considered as, in the limit that the current masses
of the up and down quarks vanish, the Goldstone bosons of a spontaneous
breaking down of the above mentioned symmetry induced by a non--vanishing
vacuum expectation value of the quark field bilinear operator $\bar\psi\psi$,
namely $\mathopen{\langle 0\,|}\bar\psi\psi\mathclose{|\, 0\rangle} \ne 0$.
Equipped with new information on
the momentum transfer dependence of one of the axial vector
form factors $g_A$ \cite{GUERRA},
we provide a refined analysis of the old results, which are based on an
extrapolation of the physical quantities
to $q^2\to 0$ \cite{AdlerBook}, to include a wider range of momentum transfer,
and, may be more interestingly, to explore possible extensions that are
observable and are nevertheless consistent with our present knowledge.

    The possibility of the formation of a superconducting phase in a massless
fermionic system, which is a realization of one of the possible phases in which
the chiral symmetry is spontaneously broken down to an isospin symmetry,
is investigated in Ref. \cite{YING1} base on a 4-fermion interaction
model. It is argued that the relativistic superconducting phase might be
relevant to the creation of baryons in the early universe due to its quantum
mechanical nature. If this scenario reflects the nature at least at a
qualitative level, it would be of interest to study the possible existence of
such a phase in the hadronic system at the present day condition.
Since it is unlikely that the present day strong interaction
vacuum at large scale is in the superconducting phase for reasons given in the
following, the only regions where the superconducting phase can be found are
inside a nucleon, a nucleus, the center region of a heavy ion collision,
and an astronomical object like a neutron star, a
quasar, etc. Albeit the possible superconducting phase in the baryon creation
era of the early universe may have had disappeared entirely at certain
previous time during the evolution of the universe, it is still a worthwhile
effort to search for such a phase at the present time. One
of the reasons is that if such a phase can be found, its properties can be
studied in domestic laboratories. Its existence is a reasonable possibility
since it is shown \cite{YING1} that
given suitable coupling constants, the Nambu Jona-Lasinio (NJL) phase in
which $\mathopen{\langle 0\,|}\bar\psi\psi\mathclose{|\, 0\rangle} \ne 0$
changes into a superconducting phase as the baryon (or quark) density is
raised. The empirical need for such an assumption will be discussed in
the conclusion parts of the paper rather than in this section since most of
the individual phenomenon considered has its own explanation in terms of
conventional picture with various degrees of success. We are interested in a
search for a consistent explanation of the observatoins. Motivated by
these considerations, we explore
some of the phenomenological consequences of a possibility in which the
interior of a nucleon contains diquark condensation \cite{diquark}, which is
enhanced in a nucleus, that spontaneously breaks the chiral symmetry.

    The model dependency of the discussion is reduced as much as possible. For
that purpose, we classify the spontaneous chiral symmetry belonging to the
same chain, namely $SU(2)_L\times SU(2)_R\to SU(2)_V$, into categories
characterized by their order parameters introduced in Ref.
\cite{YING1}, which are generic in nature. The NJL
phase is characterized by a non-vanishing $\sigma$ and vanishing
$\phi^{c\mu}$ and $\bar \phi^\mu_c$. The superconducting phase is
characterized by non-vanishing $\phi^{c\mu}$ and $\bar\phi^{\mu}_c$ and
possibly a non-vanishing $\sigma$. The generality
of the discussion allows the results to  be applied to any model with
a similar phase structure as the one obtained in Ref. \cite{YING1}.

    The paper is organized in the following way. In section 2, a
chiral  Ward-Takahashi identity is studied by presenting the main results.
Section 3 deals with the perturbation of a small
current quark mass term in the Lagrangian. Under the assumption that there is
a diquark condensation inside a nucleon, we investigate in section 4 the
necessity and the consequences of a modification of the PCAC relation and
related current algebra results. A comparison with experimental data is made.
Section 5 contains a summary and an outlook.

\section{The chiral $SU(2)_L\times SU(2)_R$ Ward-Takahashi identity}

The axial vector current vertex $iA^{5a}_\mu(p',p)$ between single
quark states is written as
\begin{eqnarray}
 iA^{5a}_\mu(p+{q\over 2},p-{q\over 2}) &=& {i\over 4}
\gamma_\mu\gamma^5\tau^a O_3 + \Gamma^{5a}_\mu(p+{q\over 2},p-{q\over
2}),\label{WT1}
\end{eqnarray}
where the initial and final state (8-component) quark spinors are
suppressed and $q_\mu$ stands for the 4-momentum transfer. Here $O_3$
and $O_{(\pm)}$ in the following are Pauli matrices acting on the
upper and lower 4-components of the 8-component spinor
$\Psi$ \cite{YING1} and $\tau^a$ $(a=1,2,3)$ is one of the Pauli
matrices acting on the flavor indices of $\Psi$. The radiative part of
the axial vector current vertex $\Gamma^{5a}_\mu$ satisfies the chiral
Ward-Takahashi identity
\begin{eqnarray}
q^\mu\Gamma^{5a}_\mu(p+{q\over 2},p-{q\over 2}) = -{i\over 4}\left (
\Sigma\gamma^5\tau^a O_3+\gamma^5\tau^a O_3\Sigma\right ),\label{WT2}
\end{eqnarray}
where $\Sigma = \sigma - \gamma\cdot\phi^c\gamma^5 {\cal A}_c O_{(+)} +
\gamma\cdot \bar\phi_c\gamma^5 {\cal A}^c O_{(-)}$ is the self-energy
term (without the contribution to the wave function renormalization)
for the quarks and ${\cal A}_{ab}^c = - {\cal A}_{c,ab} = -\epsilon^{abc}$.
$\epsilon^{abc}$ is the total antisymmetric Levi-Civita tensor in the
color space of the quark. In both the NJL phase where $\sigma \neq 0$
and $\phi^c_\mu = \bar\phi_{c\mu} = 0$ and the superconducting phase
where $\phi^c_\mu\neq 0$, $\bar\phi_{c\mu}\neq 0$ and possibly $\sigma
\neq 0$, Eq.\ref{WT2} implies that $\Gamma^{5a}_\mu$ contains a
massless Goldstone boson pole due to the fact that its right hand side
(r.h.s.) is finite in the $q^\mu\to 0$ limit. The appearance of this
massless pole in the physical excitation spectrum following the
spontaneous chiral symmetry breaking is required  by the Goldstone theorem.

 The chiral Ward-Takahashi identity Eq.\ref{WT2} can determine various
properties of the chiral Goldstone boson. We shall consider the case
in which  $\phi^2 \equiv \bar \phi_{c\mu}\phi^{c\mu} \neq 0$,
$\mu^\alpha = 0$ and $\sigma = 0$ that has not been discussed in the
literature to demonstrate some of the elementary features of the
superconducting phase related to the chiral symmetry. Eq.\ref{WT2}
becomes
\begin{eqnarray}
q^\mu\Gamma^{5a}_\mu(p+{q\over 2},p-{q\over 2}) &=& - {i\over 2} \left
( \gamma\cdot \bar\phi_c {\cal A}^c O_{(-)} + \gamma\cdot\phi^c{\cal
A}_c O_{(+)}\right ) \tau^a\label{WT3}
\end{eqnarray}
in such a case. The propagator of the Goldstone diquark is defined as
$G_\delta(q) = -i t^{\mu\nu}/\Delta(q)$. The denominator $\Delta(q)$
can be generally parameterized as
\begin{eqnarray}
\Delta(q) = q^2 + a_\delta{(\phi\cdot q)^2\over\phi^2}\label{DELTA}
\end{eqnarray}
if we choose the phases of $\phi^c_\mu$ and
$\bar\phi_{c\mu}$ such that $\phi^c_\mu = - \bar\phi_{c\mu}$. With the
following ansatz, namely
\begin{eqnarray}
t^{c'\mu\nu}_c(q\to 0) = -{{\phi^{c'\mu}\bar\phi^\nu_c}\over{\phi^2}},
&&
\bar t^{\mu c\nu}_{c'}(q\to 0) = -{\bar
\phi{^\mu}_{c'}\phi^{c\nu}\over{\phi^2}},
\label{PROJ1}
\end{eqnarray}
and with the definition of the Goldstone diquark-quark vertices given
by
\begin{eqnarray}
\bar D^{ca}_\mu(p+{q\over 2},p-{q \over 2}) = - {i\over 2} g_{\delta
  q}\gamma_\mu \tau^a {\cal  A}^c O_{(-)}, &&
D^{a}_{c\mu}(p+{q\over 2},p-{q\over 2}) = {i\over 2} g_{\delta q}
\gamma_\mu \tau^a {\cal A}_c O_{(+)},\label{DS1}
\end{eqnarray}
we can obtain the value of $a_\delta$, the Goldstone diquark-quark coupling
constant $g_{\delta
  q}$ and the Goldstone diquark decay constant $f_\delta$ by
separating out the massless pole in
$\Gamma^{5a}_\mu$, which can be approximately evaluated by using an
one loop perturbation calculation \cite{YING2}. The result is
\begin{eqnarray}
\left (\Gamma^{5a}_\mu \right )_{pole} = -2 D^b_a\left ( p+{q\over 2},p-{q
\over 2}\right ) {-it^{\alpha\beta}\over {\Delta(q)}} Tr\int {d^4k\over
(2\pi)^4} \bar D^b_\beta\left (k-{q\over 2},k+{q\over 2}\right
)\nonumber \\
{i\over \gamma\cdot \left (k+{q\over 2}\right )-\Sigma}{i\over
2}\gamma_\mu\gamma^5{\tau^a\over 2} O_3 {i\over \gamma\cdot \left
(k-{q\over 2}\right)-\Sigma} - H.c.,\label{WTpole}
\end{eqnarray}
where $H.c.$ stands for the hermitian conjugation, $Tr$ stands for the
trace operation in the Dirac, flavor, color and upper and lower
4-component spaces of $\Psi$ and the color indices are suppressed. It
can be noticed that the symmetry factor \cite{YING2} for the Feynman
diagram is different from the 4-component theory for fermions. After
a lengthy process of evaluating the trace and performing the
4-momentum integration, the above equation together with Eq.\ref{WT3}
determine the values of
$a_\delta$, $g_{\delta q}$ and $f_\delta$ \cite{YING2} as functions of
$\phi^2$. The numerical values for them are shown in Figs. 1-3. A
Goldberger-Treiman relation for single quarks in the superconducting
phase exist; it can be expressed as $g_{\delta q} f_{\delta} =
\sqrt{\phi^2}$, which also defines the scale of $f_\delta$.

\section{Small current quark mass perturbation}

A finite current quark mass that explicitly breaks the chiral $SU(2)_L\times
SU(2)_R$ symmetry has non--trivial physical consequences.
For simplicity, we shall assume that both the up and down
current quarks have an identical mass $m_0$. Some of the consequences of a
finite mass for light quarks can be  studied  base on the Ward-Takahashi
identity given by Eq.
\ref{WT2} with the term corresponding to the divergence of the axial
vector current operator taken into account, namely,
\begin{eqnarray}
q^\mu \Gamma^{5a}_\mu\left (p+{q\over 2},p-{q\over 2} \right ) &=&
{i\over 2} m_0 F_\pi \gamma^5\tau^a O_3 - {i\over 4}\left (
\Sigma\gamma^5\tau^a O_3 + \gamma^5\tau^a O_3\Sigma  \right ),
\label{WTmass1}
\end{eqnarray}
and
\begin{eqnarray}
{i\over 4}F_\pi \gamma^5 \tau^a O_3 &=& \int d^4x_1 d^4x_2
e^{ix_1\cdot(p+{q/2})-ix_2\cdot (p-q/2)}\mathopen{\langle 0\, |}
T\Psi(x_1)\bar\Psi(x_2) j^{5a}(0)\mathclose{|\,
0\rangle}|_{amp}.\label{WTmass2}
\end{eqnarray}
Here ``T'' stands for the time ordering, $j^{5a}= {i\over
4}\bar\Psi\gamma^5\tau^a O_3\Psi$ and the subscript ``amp'' denotes
the amputation of external fermion lines. $F_\pi$ is a scalar
function.

In the NJL phase,  if the assumption that $F_\pi(q^2=0)$ is  dominated by the
pion pole, the mass of the pion moves  to
a finite value,  provided that to the first order in $m_0$, the
Goldberger-Treiman relation $g_{\pi q} f_\pi  = \sigma$ and the
Gell-Mann, Oakes and Renner \cite{GOR} (GOR) relation $f^2_\pi m^2_\pi
=-{1\over 2}m_0\mathopen{\langle 0\,|} \bar \Psi \Psi \mathclose{|\, 0\rangle}$
hold. This can be
checked by evaluating the r.h.s. of Eq.\ref{WTmass1}. The $\Sigma$
term on the r.h.s. of Eq.\ref{WTmass1} has a simple diagonal form
\begin{eqnarray}
\Sigma &=& \left (\begin{array}{cc}
                   \sigma & 0\\
                    0 & \sigma
                  \end{array}
           \right )\label{SigMat}
\end{eqnarray}
in the NJL phase. It can be shown that Eq.\ref{WTmass1} is \cite{TEXTB}
\begin{eqnarray}
q^\mu \Gamma^{5a}_\mu\left (p+{q\over 2},p-{q\over 2} \right ) &=&
{i\over 2}\left ({m_0\over 2} {g_{\pi q}\over f_\pi\sigma} D_\pi(q^2)
\mathopen{\langle 0\,|} \bar\Psi\Psi\mathclose{|\, 0\rangle}-1\right
)\gamma^5\tau^a O_3,\label{NJLAdiv}
\end{eqnarray}
where
\begin{eqnarray}
D_\pi(q^2) = {1\over q^2-m_\pi^2} + \bar D(q^2).\label{Dpi}
\end{eqnarray}
$\bar D(q^2)$ is a smooth function of $q^2$ at small $q^2$ (namely,
$q^2 \leq m_\pi^2$). If the above mentioned Goldberger-Treiman
relation and the GOR relation hold and $\bar D(q^2=0) << 1/m_\pi^2$,
the r.h.s. of Eq.\ref{NJLAdiv} vanishes at $q^2=0$. Therefore
$\Gamma^{5a}_\mu$ is regular at $q^2 = 0$. It implies the
disappearance of the massless pole in $\Gamma^{5a}_\mu$ as well as in
the physical spectrum.

In the superconducting phase where $\phi^2\neq 0$ and possibly
$\sigma\neq 0$, the situation is more complicated. In this case,
$\Sigma$ term on the r.h.s. of Eq.\ref{WTmass1} takes the following
form
\begin{eqnarray}
\Sigma &=& \left (\begin{array}{cc}
            \sigma & - \gamma\cdot \phi^c\gamma^5 {\cal A}_c\\
            \gamma\cdot\bar\phi_c\gamma^5 {\cal A}^c & \sigma
                  \end{array}
           \right ).
\end{eqnarray}
A finite $m_0$ for
the current quarks in this case does not render the r.h.s. of Eq.
\ref{WTmass1} vanish when $q^\mu\to 0$. There always remains a finite
strength of the massless excitation in $\Gamma^{5a}_\mu$ as long as
the mass  term is of the form ${1\over 2} m_0 \bar\Psi\Psi$. As a
consequence, there are massless excitations in the superconducting
phase even if the chiral $SU(2)_L\times SU(2)_R$ symmetry is
explicitly broken by $m_0$. There is no GOR type of relation for the
Goldstone diquark in the superconducting phase. In addition, certain
mixing between two sets of the auxiliary fields $\pi^a$ and
($\delta^{c a}_\mu,\bar\delta^a_{c\mu}$) \cite{YING1} is needed to
represent the Goldstone boson excitation in the superconducting phase
even when $m_0\neq 0$ and $\sigma = 0$. The above mentioned mixing provides
us with one of the motivations for the following extension of the PCAC
relationship.

\section{The extension of the PCAC relation and current algebra results}

The present day large scale strong interaction vacuum is expected to
be in the NJL phase. There are a few obvious reasons for this
statement. First, the overwhelming color confinement at the present
day condition prevents the large scale superconducting phase in the
strong interaction vacuum scenario from been acceptable. Second, the
long range strong interaction force in the superconducting phase due
to the massless Goldstone diquark excitation inside the hadronic
system is absent in the experimental observations. However, localized
superconducting phases inside a baryonic system are not implausible
possibilities.

Due to the above consideration and the one given in the introduction,
we explore in this section a subset of the
phenomenological consequences of the possibility in which the interior
of a nucleon contains diquark condensation that spontaneously breaks
the chiral $SU(2)_L\times SU(2)_R$ symmetry down to an isospin
$SU(2)_V$ symmetry. The study of Ref. \cite{YING3} indicates that the
model Lagrangian introduced in Ref. \cite{YING1} indeed support such a
scenario when the coupling constant $\alpha_3$ is sufficiently large.

The on shell matrix elements of the axial vector current operator
between single  nucleon states can be parameterized as
\begin{eqnarray}
\mathopen{\langle p'\,|} A^a_\mu(0)\mathclose{|\, p\rangle} &=& \bar U(p')\left
(g_A\gamma_\mu+g_P
q_\mu + g_T{i\sigma_{\mu\nu}q^\nu\over 2 m_N}\right )\gamma^5
{\tau^a\over 2} U(p), \label{AxialMatr1}
\end{eqnarray}
with $q_\mu = (p'-p)_\mu$, $m_N$ the mass of a nucleon and U(p) the
4-component nucleon spinor. The longitudinal piece $g_P q_\mu$ on the
r.h.s. of Eq. \ref{AxialMatr1} is dominated by the contribution of the
Goldstone bosons of the spontaneous chiral symmetry breaking. If only
the NJL phase is considered, $g_P$ is given by
\begin{eqnarray}
g_P(q^2) &=& - 2 {g_{\pi N}(q^2)f_\pi(q^2)\over {q^2-m_\pi^2}} + \bar
g_P(q^2),\label{GP}
\end{eqnarray}
with $\bar g_P(q^2)$ the residue term  and $g_{\pi N}(q^2)f_\pi(q^2)$
a slow varying function of both $m_\pi^2$ and $q^2$. If, however, the
assumption that there is a diquark condensation inside a nucleon is
made, there world be another longitudinal term in the matrix elements
of the axial vector current operator due to the Goldstone diquark
excitation inside  that nucleon. The expression for $g_P$ has to be
modified to
\begin{eqnarray}
g_P(q^2) &=& -2 \left ({g_{\pi N}(q^2)f_\pi(q^2)\over {q^2-m_\pi^2}} +
                      z_\delta g_{\delta
N}(q^2)f_\delta(q^2)\eta(q^2)\right ) + \bar g_P(q^2)\label{GP2}
\end{eqnarray}
after considering this additional excitation. Here $\eta(q^2)$ is
related to the propagator of the Goldstone diquark excitation and
$z_\delta$ is a constant. Similar to the pion, we introduce $g_{\delta N}$
as the Goldstone diquark-nucleon coupling constant and $f_\delta$ as
the Goldstone diquark decay constant. Albeit there is massless
excitation in $\Gamma^{5a}_\mu$, there is  no pole behavior in
$\eta(q^2)$ in  the small $q^\mu$ region due to the fact that a
Goldstone diquark  carries color so that it is confined inside the
nucleon.

The PCAC relation is given by
\begin{eqnarray}
\partial^\mu A^a_\mu &=& - f_\pi m_\pi^2 \phi_\pi^a.\label{PCAC2}
\end{eqnarray}
It can be regarded as a definition of the pion field in the low energy
regime (when going off the pion mass shell). Taking the matrix elements
of Eq. \ref{PCAC2}, it can be shown that this definition is
inconsistent with Eq.\ref{GP2} due to the additional term added to
$g_P$. We have at least two choices. The first one is to reject
Eq.\ref{GP2}, which we shall not do in this paper. The second one is
to modify Eq.\ref{PCAC2} when its matrix elements are taken. It can be
specified as
\begin{eqnarray}
  \mathopen{\langle p'\,|} \partial^\mu A^a_\mu \mathclose{|\, p\rangle} &=& -
\mathopen{\langle p'\,|}\left (
f_\pi m_\pi^2 \phi^a_\pi + f_\delta s_\delta \phi_\delta^a\right )
\mathclose{|\, p\rangle},
\label{PCAC3}
\end{eqnarray}
with $s_\delta \sim m_\pi^2$ a parameter proportional to $m_0$ and
$\phi_\delta^a$ a pseudo-scalar \cite{drive}
driving field for the Goldstone diquark excitation inside the nucleon.

The r.h.s. of Eq.\ref{PCAC3} is assumed to be dominated by the chiral
Goldstone boson contributions in the sequel. The following relations
hold, namely,
\begin{eqnarray}
g_P(q^2) &=& -2 {m_N g_A(q^2)\over q^2 - m_\pi^2},\label{PCAC4}\\
 m_N g_A(q^2) &=&
g_{\pi N}(q^2) f_\pi(q^2) + (q^2 - m_\pi^2) {s_\delta \over m_\pi^2}
g_{\delta N}(q^2) f_\delta (q^2) \eta (q^2), \label{PCAC5}
\end{eqnarray}
when we require that the coefficients of $m_\pi^2$ and $q^2$ vanish
separately while assuming that $(q^2-m_\pi^2)g_P(q^2)$ and $g_A(q^2)$
are slow varying functions of both $m_\pi^2$ and $q^2$. The slow
varying assumption of these quantities together with the modified PCAC
relation, Eq.\ref{PCAC3}, imply that the residue term $\bar g_P(q^2)$
in Eq.\ref{GP2} is unimportant in the small $q^2$ regime. Eqs.
\ref{PCAC4}, \ref{PCAC5} are consistent with Eq.\ref{GP2} provide that
$z_\delta =
s_\delta/m_\pi^2$. The modified  Goldberger-Treiman relation is
obtained when $q^2 = m_\pi^2$ is  assumed on the r.h.s. of the second
equation of Eqs.\ref{PCAC4}, \ref{PCAC5} and $q^2\to 0$ is taken on its left
hand
side (l.h.s.), namely,
\begin{eqnarray}
m_N g_A(0) &=& g_{\pi N}(m_\pi^2)f_\pi(m_\pi^2) + \lim_{q^2\to m_\pi^2}
(q^2-m_\pi^2)z_\delta g_{\delta N}(q^2) f_\delta(q^2)\eta(q^2).\label{GP3}
\end{eqnarray}
Assuming the results of Ref. \cite{GUERRA} can be used in the time like
region for $q_\mu$, the extrapolation of $g_A$ from $q^2 = m_\pi^2$ to
$q^2=0$ introduces an error of order $m_\pi^2/M_A^2\sim 1 \%$ with $M_A
\sim 1 GeV$ \cite{GUERRA}.

The Goldberger-Treiman relation is satisfied to about $5\%$ in
experimental observations. This good agreement indicates  the validity
of the various assumptions combined made above and a rather small
value of $\lim_{q^2\to m_\pi^2} (q^2-m_\pi^2)z_\delta g_{\delta
N}(q^2)f_\delta(q^2)\eta(q^2)$. The $\eta(q^2)$ term on the pion mass
shell actually  vanishes since $\eta(q^2)$ does not has a pole in the
low $q^\mu$ region. However, a small value for the $\eta(q^2)$ term
does not  follow from the success  of the Goldberger-Treiman relation.

The effects of the Goldstone diquark inside a nucleon, if exist,
might be comparatively large in the kinematic region off the pion
mass shell. Current experimental data considered do not allow any
conclusive statement to be made on this point. The value of $g_P$
can be compared to the PCAC  value (Eq.\ref{GP}) to detect the
possible effects of the Goldstone diquark. Experimental determinations
of $g_P$ for a nucleon in muon capture experiments involving
light nuclei show a systematic increase (of order as large as $100\%$)
of the value of $g_P$  from its PCAC one \cite{TOWNER}. The world
average value of $g_P$ obtained from the muon capture experiments on
a hydrogen is close to the one given by the PCAC one \cite{BARDIN1}.
However, the interpretation of the results is not unambiguous
(Ref. \cite{BARDIN1} and Gmitro and Truol in Ref. \cite{TOWNER}).
There are two
recent measurements of the muon capture rate on the deuterium. The
central value of the first measurement \cite{BARDIN2} implies a value of
$g_P$ smaller than the PCAC one \cite{TATARA}. The central value of the
second measurement \cite{CARG} implies a value of $g_P$ close to the
PCAC one \cite{TATARA}. The problems related to the value of $g_P$ are not
yet completely settled \cite{RECENT}.
Polarized $\beta$ decay experiments offer
alternative means of measuring the value of $g_P$ in a different range
of $q^2$ \cite{YING4}. The experiments are difficult but in principle
possible. The $q^2$ dependence of $g_A(q^2)$ and $g_{\pi N}(q^2)
f_\pi(q^2)$ differs from each other (see Eqs.\ref{PCAC4}, \ref{PCAC5}) if
there are Goldstone diquark excitations inside a nucleon. More and
better experimental data are needed to investigate such a deviation.

Theoretically, the agreement of the PCAC value for $g_P$ and the
experimentally measured one is expected for a nucleon in free space.
This is because Eq.\ref{PCAC4} implies the deviation of $g_P$ from the
PCAC one is related to the deviation of the value of $g_A$, $m_N$ and $m_\pi$
from the experimentally observed ones, which is not true. Indeed, a
recent experimental study \cite{choi} provides a strong support \cite{explain}
of Eq. \ref{PCAC4} within momentum transfer of $0<-q^2 < 0.2\hspace{3pt}
GeV^2$. Therefore,
the system in which a deviation of the strength of longitudinal
modes in the axial vector current operator from the PCAC
one can be observed is inside a nucleus with relatively large number
of nucleons. In these systems, the coupling between the pions outside
of the nucleons and the Goldstone diquark excitations inside the
nucleons could render the properties of pions to change to such a degree that
a deviation from PCAC value can be observed.

The modification of some of the current algebra results  related to
PCAC due to a possible Goldstone diquark excitation inside a nucleon
can also be investigated. The Adler-Weisberger sum rule    and the
nucleon $\Sigma_N$ term will be studied in this paper base on a Ward
identity involving the axial vector current operator \cite{ADLER,WEISB,BPP}.
In order to obtain useful  information, the low lying longitudinal excitation
contributions and the rest part of the axial vector current operator
inside a time ordered product are separated in the following way
\begin{eqnarray}
\left <T\left (\ldots A_\mu^a \ldots \right )\right > &=& \left <
T\left (\ldots\bar A_\mu^a\ldots\right )\right > +\nonumber\\
&& \partial_\mu
\left < T\left (\ldots f_\pi\phi_\pi^a\ldots \right )\right > +
\partial_\mu\left < T\left (\ldots z_\delta
f_\delta\phi_\delta^a\ldots\right )\right >, \label{Adecomp}
\end{eqnarray}
with the second and the third terms on the r.h.s. the longitudinal
parts of $A_\mu^a$, which is dominated by the low lying chiral
Goldstone boson contributions, and $\bar A_\mu^a$, which is expected
to change slowly with the momentum transfer $q^\mu$ when its matrix
elements are taken between nucleon states, containing the rest part of
$A_\mu^a$. The matrix elements are evaluated on the pion mass shell
\cite{BPP}. Those of $\bar A_\mu^a$ that connect nucleon states by
a gradient--coupling \cite{Weinberg} are
then extrapolated to the kinematic point where $q^2=0$ in order to
compare with experimental data by taking the advantage that they are
slow varying function of $q^2$. The  error of the extrapolation is
expected to be of order $O(m_\pi^2/M_A^2)\sim 1\%$. There are
contributions from other off (nucleon) shell terms and baryonic
excitations in the intermediate states; they are not investigated in
detail here. The modified Adler-Weisberger sum rule can be shown to
have the following form
\begin{eqnarray}
g_A^2(q^2=0)&=&1-2{f_\pi^2(m_\pi^2)\over \pi}\int_{m_\pi}^\infty d\nu
{\sigma^{\pi^-p}_{tot} (\nu)-\sigma^{\pi^+ p}_{tot}(\nu)\over
  (\nu^2-m_\pi^2)^{1/2}}
\nonumber\\
&& - 2\lim_{q^2\to m_\pi^2}(q^2-m_\pi^2)^2 f_\delta^2(q^2)\lim_{\nu
  \to 0} {z_\delta^2\over\nu} G_{\delta N}^{(-)}(\nu,0,q^2,q^2),\label{gA2}
\end{eqnarray}
where $G_{\delta N}^{(-)}$ is related to the isospin odd forward
Goldstone diquark-nucleon scattering amplitude, which is driven by
$\phi_\delta^a$, without the amputation of the diquark lines.

The value of $g_A$ from nuclear $\beta$ decay experiments is $1.261\pm
0.004$ \cite{PREV}. The value of the same quantity obtained from the
Adler-Weisberger sum rule using pion nucleon scattering cross section
(and  after estimating other corrections) is about $1.24\pm 0.02$
 \cite{ADLER}. These two numbers, with a central ($g_A-1$) value
deviation of order $10\%$, barely overlap. Again, the contribution of
the second term on the r.h.s. of Eq. \ref{gA2} is expected to be zero
on the pion mass shell. Therefore the good agreement of the
Adler-Weisberger sum rule with the observation does not necessarily
constitute a fact that is against the additional terms added in Eqs.
\ref{PCAC3} and \ref{Adecomp}.

The expression for the nucleon $\Sigma_N$ term, which is of order
$m_0$, is modified to
\begin{eqnarray}
\Sigma_N &=& f_\pi^2(m_\pi^2) T^{(+)}_{\pi N}(0,0,m_\pi^2,m_\pi^2) +
\lim_{q^2\to m_\pi^2} (q^2-m_\pi^2)^2 f_\delta^2(q^2)z_\delta^2 G_{\delta
  N}^{(+)}(0,0,q^2,q^2),\label{SigmaN}
\end{eqnarray}
with $G^{(+)}_{\delta N}$ related to the isospin even forward
Goldstone diquark-nucleon scattering amplitude without the amputation
of the diquark lines. The second term on the r.h.s. of Eq.\ref{SigmaN}
vanishes on the pion mass shell. Experimental verifications of
Eq.\ref{SigmaN} (without the second term on its r.h.s.) have so far
been unsuccessful without assuming certain strange quark content of a
nucleon, which is discouraged by the OZI rule \cite{OZI}. A recent
review of the nucleon $\Sigma_N$ problem can be found in Ref.
 \cite{GLS}. There   is still a sizable discrepancy \cite{GLS} of order
$25\%$ between the one obtained from the pion-nucleon scattering data
($45 MeV$) and the one obtained from  other available sources ($35
MeV$). The problem is still not well understood.

It should be pointed out here that an additional source of correction
need to be considered to relate $\Sigma_N$, which is proportional to
the scalar density of the nucleon on the Cheng-Dashen point \cite{GLS},
to $\sigma_N$ related to the scalar density of a nucleon. In the
Hartree-Fock approximation, the static scalar density measured by
$\sigma$ is free of radiative corrections due to the fact that it
satisfies a ``gap equation'', which self-consistently adjust the
radiative corrections to $\sigma$ to zero (cancel) by changing its
value. This can be proven in the case that $\sigma$ is space-time
independent \cite{YING2}. We expect it to be true also for $\sigma_N$.
This statement is not true if the matrix elements of the scalar
density operator ${1\over 2}\bar\Psi\Psi$ between states of different
4-momentum are taken. Various corrections due to the interaction have
to be considered.

$\Sigma_N = \sigma_N(m_\pi^2)$ term is written as $\Sigma_N =
\sigma_N(0) + \Delta_N$, $\sigma_N(0)=25 MeV$ can be obtained from the
baryon spectrum \cite{GLS}. $\Delta_N$ contains various corrections to
the extrapolation. If there is diquark condensation in a nucleon,
there would be contributions to $\Delta_N$ that are proportional to
$\phi^2$ coming from the interaction terms. The existence of such a
term can be demonstrated at the quark level by considering a second
order correction to $\sigma_N$ due to the interaction, which has the
generic form $\delta \sigma_N^{(2)}\sim \mathopen{\langle p'\,|} T\left
(\Psi\bar
\Psi\right )\left (\bar\Psi \Gamma\Psi \right )\left (\bar\Psi \bar
\Gamma \Psi\right )\mathclose{|\, p\rangle}$, with $\Gamma$ and $\bar \Gamma$ a
pair of
the interaction vertices. This term contains $\phi^2$ contributions if
there is diquark condensation inside a nucleon \cite{diquark2}.
Detailed evaluation of these terms can not be proceeded
before a specific model is given. Here, we simply parameterize the
effects  of $\phi^2$ as
$\Sigma_N=\sigma_N(0)+\Delta_N^{(0)}+\beta_N\phi_N^2$, where
$\Delta_N^{(0)}$ is related to the total correction that has already
been considered in the literature (see, e.g., Ref. \cite{GLS}). The new
term depending on $\phi^2_N$ is due to the possible diquark
condensation inside a nucleon with an average strength measured by
$\sqrt{\phi_N^2}$. It is unclear at present whether the additional
term increases ($\beta_N<0$), decreases ($\beta_N>0$) or even
eliminates the above mentioned discrepancy.

\section{Summary and Outlook}

    A natural way of extending the PCAC relationship beyond the conventional
one is presented in this study. For a relativistic fermionic system, the
extension presented here is unique if one assumes that 1) in the low momentum
transfer region ($|q^2| << M^2_A \approx 1\hspace{3pt} GeV$) the matrix element
of the axial vector current operator is dominated by the collective excitations
(quasi--Goldstone bosons) related to the spontaneous broking down of the
(approximate) chiral $SU(2)_L\times SU(2)_R$ symmetry and 2) the contribution
of those less collective hadronic excitations, which have masses of order
1 $GeV$ or larger, are relatively small.

    From the analysis given above, it can be seen that the effects of a
possible Goldstone diquark mode inside a nucleon are suppressed in the
kinematic regime where pion is on its mass shell. The effects of such an
excitation mode, if exist, could be revealed in the kinematic region off
the pion mass shell, where the simple pion pole behavior of certain
observables is modified in such a way that can not be explained by
conventional pion pole saturation assumption for the divergence of the
axial vector current operator. They can be observed in, e.g., high precision
$\beta$ decay \cite{YING4,HOLSTEIN}, $\mu$ capture experiments
\cite{TOWNER,TATARA,RECENT} and in experiments that can compare the
momentum transfer dependencies of $g_A$, which is largely known,
and $g_{\pi N N} f_\pi$ (see Eq. \ref{PCAC5}).

    For a large nucleus, in which the nucleons are close together, the
effects of the diquark condensation and Goldstone diquarks are expected to
be enhanced. The enhancement of the diquark condensation can be attributed
to the increase of the volume and baryon density of the system;
the enhancement of the effects of the Goldstone diquarks is due to their
coupling to the pions outside the nucleons. These effects could shed some
light on the question \cite{BFS} of missing pions, quark number depletion
in a nucleus and violation of Gottfried sum rule \cite {GOTT} in a nucleon.
Due to the novel nature of the superconducting phenomena in a relativistic
massless fermion system, a quantitative study depends first on a model for a
nucleon, which is still under investigation, we provide a qualitative picture
in the following since if we make the assumption that there is a diquark
condensation in a nucleon which is enhanced in a nucleus then the
general trend of these phenomena follow quite naturally. Detailed study will
be given elsewhere.

   The lack of polarization effects observed in Refs. \cite{Carey} and
\cite{McClelland} can be explained, together with other many body effects, by
a reduction of pionic effects from its vacuum ones due to an
enhancement of contribution of the second term on the r.h.s. of Eq. \ref{PCAC5}
resulting from the increased Goldstone diquark excitation strength inside a
nucleus. The missing polarization phenomenon observed in pion nucleus
scattering
process is due to cancellation between various components of nucleon--nucleus
scattering \cite{GEB},
it is premature to make any definite statement at this point.
It is the question of overall consistency rather than fitting of a single
experimental observation that prevent us from doing so here.

   The depletion of quark number can be related to the missing or spreading
of the isoscalar part of the electrical charge of the partons in the
superconducting phase discussed in this paper \cite{YING2}. This phenomenon
is known
in the study of superconducting condensed matter system
\cite{Nambu}.  In such a case, the electric charge of some \cite{YING2} of the
partons, which are the up and down quarks, have only isovector charge
$\pm 1/2$ in the limit of
large strength of diquark condensation. Therefore if there is a diquark
condensation that generates a localized
superconducting phase in a nucleon, the $F_2(x)$
structure function should be written as
\begin{eqnarray}
   F_2(x) & = & \sum_i \tilde Q^2_i x f_i(x), \label{F_2equation}
\end{eqnarray}
with
\begin{eqnarray}
   \tilde Q_i & = & \alpha_i {1\over 6} \pm {1\over 2}. \label{e-charges}
\end{eqnarray}
here $0 \le \alpha_i \le 1$ characterizes the strength of the superconducting
phase. In case of strong superconductivity, which means some of the
$\alpha_i = 0$, $F_2^N(x) \sim F_2^P(x)$. The experimental measurement of
$F_2^P(x)$ and $F_2^N(x)$ in the deep inelastic scattering experiments
shows such a tendency manifests in the violation of Gottfried sum rule
\cite{GOTT}. One of the interpretation is that there is
an isospin violation in the
sea \cite{ISOv} due to pionic collective excitation. If we use the normal
electrical charges for the up and down quark,
then there is a reduction of $f_i(x)$, which is interpreted as a violation of
the Gottfried sum rule.

In a large nucleus, the effects of diquark condensation are expected to be
enhanced; this in general produces the depletion of quark number observed
in the EMC effects in a nucleus, which is related to a (enhanced) violation
of Gottfried
sum rule in the picture given here,  since the antiquark
components in a nucleus is known to be small in experimental observations
\cite{Eisele,Anti-quark}. It is generally accepted that the EMC effects are
still not well understood \cite{BPT}.

A detailed study of the effects of these soft longitudinal modes and that of
spreading of the isoscalar part of the charge of constituent quarks
is beyond the scope of this paper. The above study shows that a coherent
picture seems to be emerging if the assumption of a diquark condensation
is made. It is an interesting topic to be explored.

\section*{Acknowledgment}

  The author would like to thank the Institute for Nuclear Theory at
the University of Washington for its hospitality during the completion of
the work.

\section*{Figure Captions}

\noindent
Figure~1: The  $\phi^2$ dependence of $a_\delta$.  $\Lambda$ is the
chiral symmetry breaking scale.
In order to obtain a smooth curve, the sharp cut off  in the Euclidean
momentum space integration is replaced by a smooth one. The functional
dependence of the smooth cut off $F(p/\sqrt{\phi^2})$ is plotted on
the same graph. Since $a_\delta > 1 $, a Goldstone diquark can only
propagates outside  of the cone $\theta_0 = cos^{-1}\sqrt{1/a_\delta}$
in the direction of ${\vec{\phi}}$ in the frame where $\phi^0 = 0$.

\vskip\baselineskip
\noindent
Figure~2: The  $\phi^2$ dependence of the quark-diquark coupling
constant $g_{\delta q}$. The
same smooth Euclidean momentum space cut off as the one used in Fig. 1 is used.

\vskip\baselineskip
\noindent
Figure~3: The  $\phi^2$ dependence of the Goldstone diquark decay
constant $f_\delta$. The same smooth Euclidean momentum space cut off
as the one used in Fig. 1 is used.


\begin{thebibliography}{99}

\bibitem{GUERRA}  A. del Guerra et al., Nucl. Phys. {\bf B107} (1976) 65; W.
A. Mann et al., Phys. Rev. Lett. {\bf 31} (1973) 844.
\bibitem{AdlerBook} See, e.g., S. L. Adler and R. F. Dashen, {\it
                   Current Algebra and Applications to Particle Physics}
                   (W. A. Benjamin, Inc., New York, Amsterdam, 1968).
\bibitem{YING1}  S. Ying, Phys. Lett. {\bf B283} (1992) 341.
\bibitem{diquark} The condensation of diquark can be defined by a localized
                  stablizing non--vanishing $\phi_{c\mu}$ and
                  $\bar\phi^{\mu}_c$ auxiliary fields, which are introduced in
                  Ref. \cite{YING1}, inside the nucleon. By definition, all
                  quantum fluctuation in the quantum fields are included.
\bibitem{YING2}  S. Ying, in preparation.
\bibitem{GOR}  M. Gell-Mann, R. Oakes, and B. Renner, Phys. Rev.
{\bf 175} (1968) 2195; R. Dashen, Phys. Rev. {\bf 183} (1969) 1245.
\bibitem{TEXTB}  For additional information, see standard text books,
    e.g., S. Pokorski, {\it Gauge Field Theories} (Cambridge University Press,
    Cambridge, 1987).
\bibitem{YING3}  S. Ying, unpublished.
\bibitem{drive} A pseudo-scalar field can excite the axial vector Goldstone
                diquark inside a nucleon due to the presence of  baryonic
                matter.
\bibitem{TOWNER}  I. S. Towner, Annu. Rev. Part. Sci. {\bf 36} (1986) 115;
S. Nozawa, K. Kubodera, and H. Ohtsubo, Nucl. Phys. {\bf A453} (1986) 645;
M. Gmitro and P. Truol, Adv. Nucl. Phys. {\bf 18} (1987) 241; W. C. Haxton
and C. Johnson, Phys. Rev. Lett. {\bf 65} (1990) 1325; D. S. Armstrong
et al., Phys. Rev. {\bf C43} (1991) 1425.
\bibitem{BARDIN1} G. Bardin et al., Nucl. Phys. {\bf A355} (1981) 365.
\bibitem{BARDIN2} G. Bardin et al., Nucl. Phys. {\bf A453} (1986) 591.
\bibitem{TATARA} N. Tatara, Y. Kohyama, and K. Kubodera, Phys. Rev.
{\bf C42} (1990) 1694;
S. Ying, W. Haxton, and E. M. Henley, Phys. Rev. {\bf C45} (1992) 1982.
\bibitem{CARG} M. Cargnelli, dissertation, Universitat Wien, 1987; M.
Cargnelli et al., in Proceedings of the XXIII Yamada Conference on
Nuclear Weak Processes and Nuclear Structure, Osaka,  1989, edited by
M. Morita, H. Ejiri, H. Ohtsubo, and T. Sato (World Scientific,
Singapore, 1989), p.115.
\bibitem{RECENT} Recent theoretical efforts are: H. W. Fearing, M. S. Welsh,
                Phys. Rev. {\bf C46} (1992) 2077;
                J. G. Congleton, H. W. Fearing, Nucl. Phys. {\bf A552}
                (1993) 534.
\bibitem{YING4} S. Ying, Phys. Rev. {\bf C47} (1993) 833.
\bibitem{choi} S. Choi, et al., Phys. Rev. Lett. {\bf 71} (1993) 3927.
\bibitem{explain} The relationship given by Eq. \ref{PCAC5} is a prediction of
                  this investigation completed at the end of 1992.
\bibitem{ADLER} S. L. Adler, Phys. Rev. {\bf 140} (1965) 736.
\bibitem{WEISB} W. I. Weisberger, Phys. Rev. {\bf 143} (1966) 1302.
\bibitem{BPP} L. S. Brown, W. J. Pardee, and R. D. Peccei, Phys. Rev.
              {\bf D4} (1971) 2801.
\bibitem{PREV} Review of Particle Properties, Phys. Lett. {\bf B239} (1990).
\bibitem{OZI} S. Okubo, Phys. Lett. {\bf 5} (1963) 165; G. Zweig, CERN
Report, No. 8419 (1964),  unpublished; I. Iuzuka, Prog. Theor. Phys. Suppl.
{\bf 37-38} (1966) 21.
\bibitem{Weinberg} S. Weinberg, Phys. Rev. Lett. {\bf 17} (1966) 168.
\bibitem{Comput} In an accurate calculation, this kind of extrapolation can
be replaced by a extrapolation of the empirically known $g_A(q^2)$
\cite{GUERRA} in the space--like region to the time--like region,
which gives, hopefully, better values for $g_A(q^2)$.
\bibitem{GLS} J. Gasser, H. Leutwyler, and M. E. Sainio, Phys. Lett.
{\bf B253} (1991) 252.
\bibitem{HOLSTEIN} See, e.g., B. R. Holstein, Rev. Mod. Phys., {\bf 46}
 (1974) 789; see also S. Ying, Ph.D. thesis (University of Washington at
Seattle, 1992), unpublished.
\bibitem{diquark2} There is a
  normal ordered operators $:\bar\Psi_\alpha\bar\Psi_\beta\Psi_\gamma
  \Psi_\sigma:$ (with respect to the NJL vacuum) in
  the Wick expansion of the time ordered product of the fermion fields
  in $\delta \sigma_N^{(2)}$. Part of their contributions to
  $\delta\sigma_N^{(2)}$, which are proportional to $\phi^2$, are
  non-vanishing if there is a diquark condensation inside a
  nucleon.
\bibitem{BFS} G. F. Bertsch, L. Frankfurt, and M. Strikman, Science,
              {\bf 259} (1993) 773.
\bibitem{GOTT} A. Amaudrus et al. (New Muon Collaboration),
               Phys. Rev. Lett. {\bf 66} (1991) 2712.
\bibitem{Carey} T. A. Carey et al., Phys. Rev. Lett. {\bf 53} (1984) 144;
                J. B. Rees et al., Phys. Rev. {\bf C34} (1986) 627.
\bibitem{McClelland} J. B. McClelland et al., Phys. Rev. Lett. {\bf 69}
                     (1992) 582.
\bibitem{GEB} G. E. Brown, M. Buballa, Z. B. Li, and J. Wambach,
               Preprint (SUNY-NTG-94-54, 1994).
\bibitem{Nambu} Y. Nambu, Phys. Rev. {\bf 117} (1960) 648.
\bibitem{ISOv} E. M. Henley and G. A. Miller, Phys. Lett. {\bf 251} (1990)
               453; A. I. Signal, A. W. Schreiber and A. W. Thomas, Mod. Phys.
               Lett. {\bf A6} (1991) 271; W. Melnitchouk, A. W. Thomas and A. I
               Signal, Z. Phys. {\bf A340} (1991) 85; S. Kumano and J. T.
               Londergan, Phys. Rev. {\bf D44} (1991) 717; A. Szczurek and J.
               Speth, Nucl. Phys. {\bf A555} (1993) 249.
\bibitem{Eisele} F. Eisele, Rep. Prog. Phys. {\bf 49} (1986) 233.
\bibitem{Anti-quark} R. G. Arnold et al., Phys. Rev. Lett. {\bf 52} (1984) 727;
                     M. Arneodo et al., Phys. Lett. {\bf B211} (1988) 493;
                     D. M. Alde et al., Phys. Rev. Lett. {\bf 64} (1990)
                     2479.
\bibitem{BPT}  R. P. Bickerstaff and A. W. Thomas, J. Phys. {\bf G15} (1989)
               1523.
\end{thebibliography}
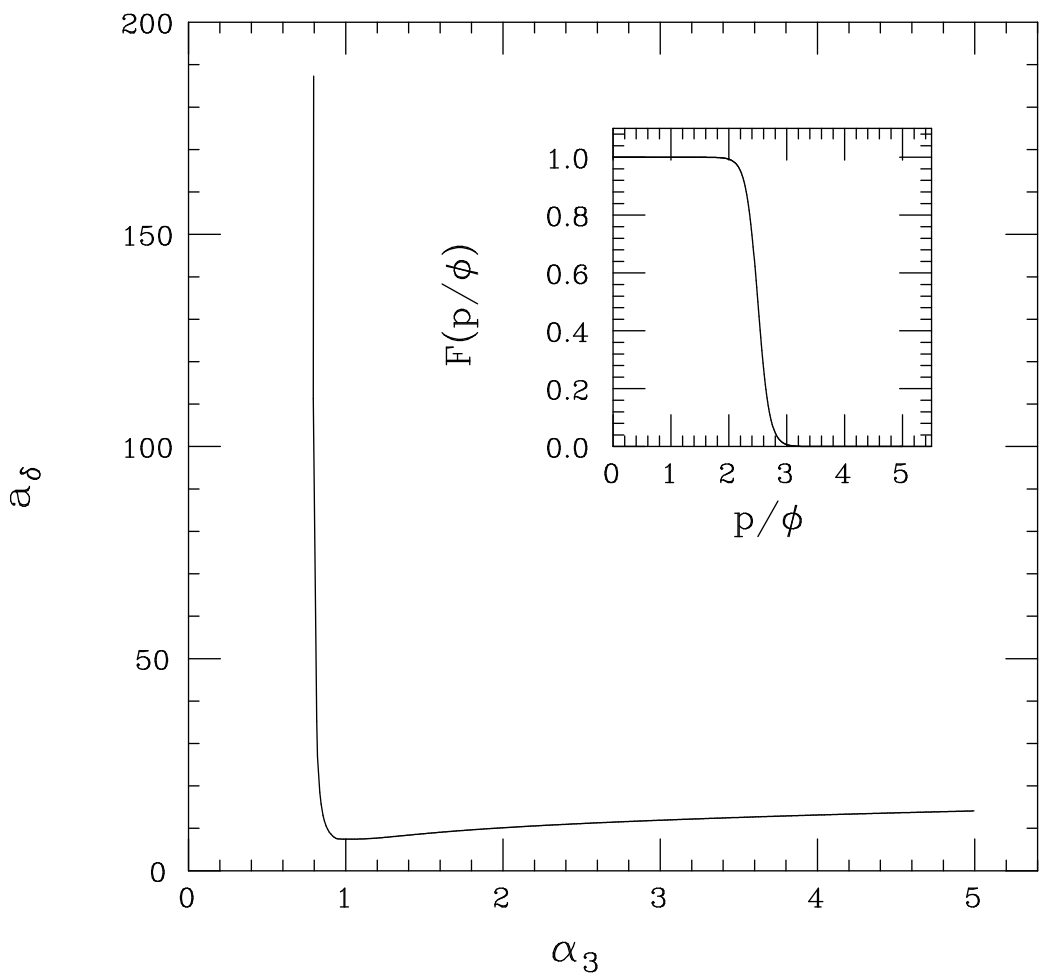
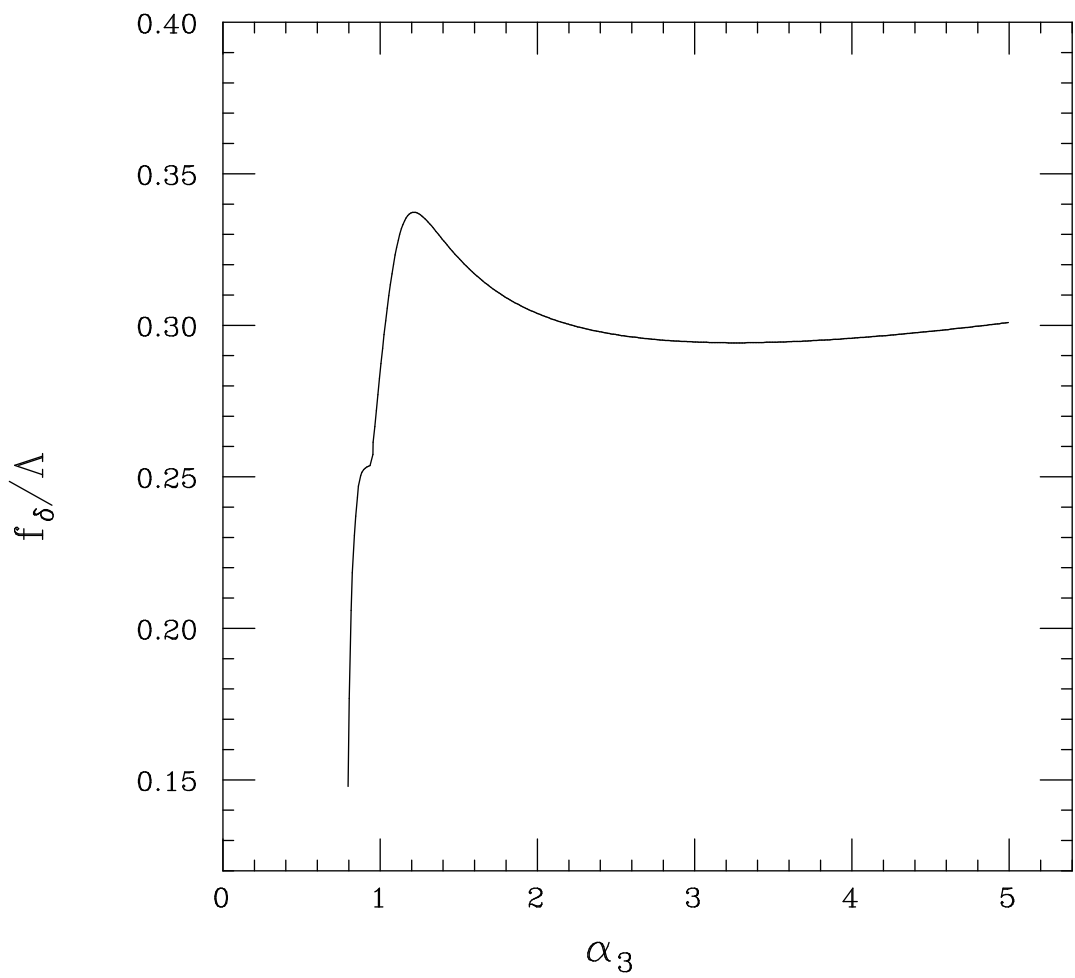
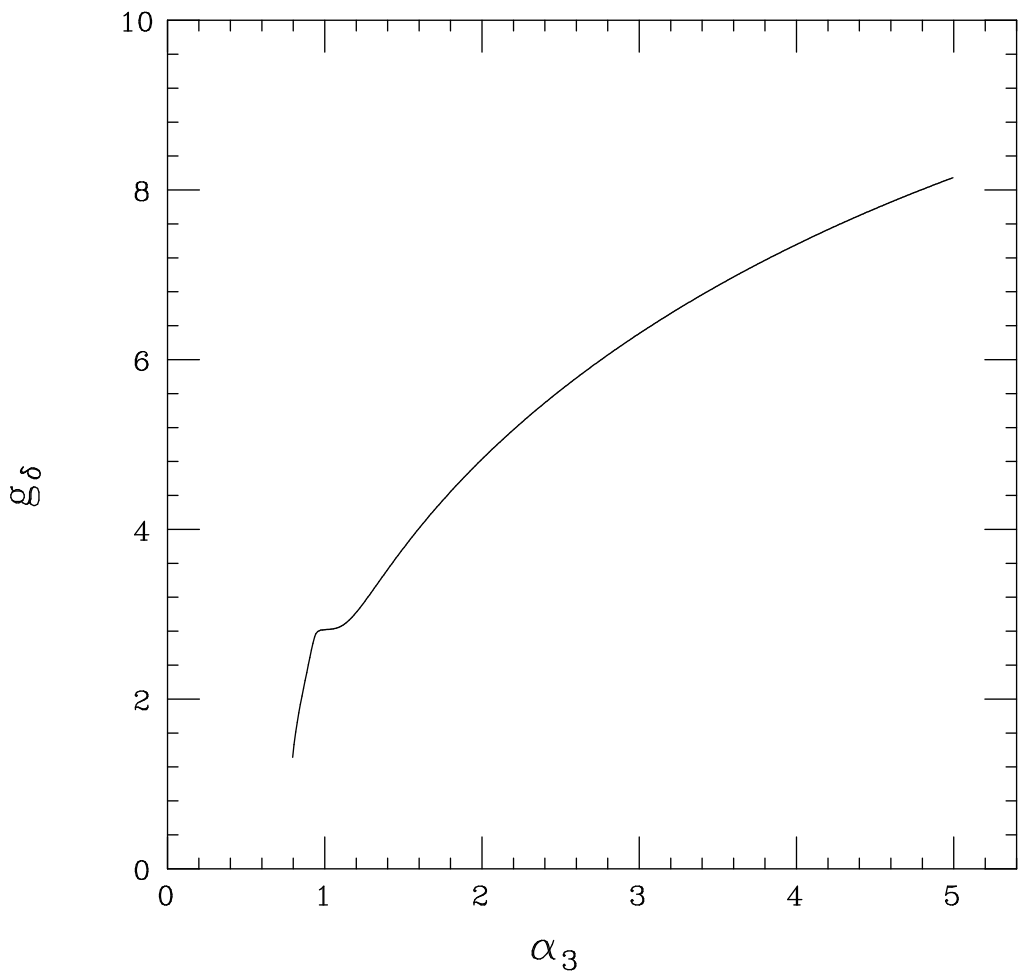
\end{document}